\documentclass[conference]{IEEEtran}
\usepackage[latin1]{inputenc}
\usepackage{times,amsmath}
\usepackage{amssymb}
\usepackage{pstool}
\usepackage{subfigure}
\usepackage{multirow}
\usepackage{enumerate}
\usepackage{graphicx}
\usepackage{MnSymbol}
\usepackage{stfloats} 
\usepackage[table]{xcolor}
\usepackage[square, comma, sort&compress, numbers]{natbib}
\usepackage{nohyperref}
\usepackage{algorithm,algorithmic}
\usepackage{epstopdf}

\graphicspath{ {Figures/} }

\begin{document}

\title{Location-based Physical Layer Authentication in Underwater Acoustic Communication Networks}

\author{
\IEEEauthorblockN{
Waqas Aman\IEEEauthorrefmark{1}, Saif Al-Kuwari\IEEEauthorrefmark{1}, Marwa Qaraqe\IEEEauthorrefmark{1} }
\IEEEauthorblockA{\IEEEauthorrefmark{1}Division of Information and Computing Technology, College of Science and Engineering, \\Hamad Bin Khalifa University, Qatar Foundation, Doha, Qatar. \\waqasaman87@gmail.com, smalkuwari@hbku.edu.qa, mqaraqe@hbku.edu.qa}
}

\maketitle

\begin{abstract} 
Research in underwater communication is rapidly becoming attractive due to its various modern applications. An efficient mechanism to secure such communication is via physical layer security. In this paper, we propose a novel physical layer authentication (PLA) mechanism in underwater acoustic communication networks where we exploit the position/location of the transmitter nodes to achieve authentication. We perform transmitter position estimation from the received signals at reference nodes deployed at fixed positions in a predefined  underwater region. We use time of arrival (ToA) estimation and derive the distribution of inherent uncertainty in the estimation. Next, we perform binary hypothesis testing on the estimated position to decide whether the transmitter node is  legitimate or malicious. We then provide closed-form expressions of false alarm rate and missed detection rate resulted from binary hypothesis testing. We validate our proposal via simulation results, which demonstrate errors' behavior against the link quality, malicious node location, and receiver operating characteristic (ROC) curves. { We also compare our results with the performance of previously proposed fingerprint mechanisms for PLA in underwater acoustic communication networks, for which we show a clear advantage of using the position as a fingerprint in PLA}.
\end{abstract}


\section{Introduction}
\label{sec:intro}
Underwater acoustic communication has gained significant attention due to its promising applications in marine life exploration, natural resource finding, underwater navigation, and military operations \cite{song2019editorial}. Recently, a considerable amount of research has been conducted to explore innovative approaches from design to signal processing techniques of underwater acoustic communication \cite{fattah2020survey}. However, the broadcast nature of underwater acoustic communication makes it vulnerable to many types of malicious attacks \cite{aman2022security}. One of the prominent attacks is the impersonation attack (also called spoofing), where a malicious node aims to mimic one of the legitimate nodes of the network in order to get access and compromise the integrity of the system  \cite{Ammar:VTC:2017,waqas:Sensors:2021}. Traditionally, such attacks were countered using pre-defined secrets (also called passwords or keys) secured via different encrypted algorithms \cite{kumari2015user}. However, recent advances in computational resources and
quantum computing may potentially jeopardize these crypto-based solutions \cite{arbaugh2002your,gidney2021factor}. Consequently, physical layer authentication (PLA), where physical layer features serve as device fingerprints. is becoming an attractive alternative approach to cryptography-based authentication. Both channel and hardware-based features can be used for PLA. Such hardware features include I/Q imbalance \cite{hao2014performance} and carrier offsets (phase and frequency) \cite{rahman2017exploiting,rahman2014phy}  while channel features include pathloss \cite{waqas:Sensors:2021}, channel impulse response (CIR) \cite{Ammar:VTC:2017}, channel frequency response \cite{xiao2008using} and received signal strength \cite{yang2012detection}. 

To counter the impersonation attack in underwater  acoustic communication networks, several works were recently published considering PLA \cite{Waqas:Access:2018,xiao2018learning,khalid2020physical,9676618}.
To the best of our knowledge, the first work exploiting distance and angle of arrival as device fingerprints for PLA in underwater acoustic sensor networks is reported in \cite{Waqas:Access:2018}, where closed-form expressions for error probabilities are derived. Similarly, the authors in \cite{xiao2018learning} study the power delay profile of underwater acoustic channels using a deep reinforcement learning approach to detect impersonation. Adopting a different approach, the authors in \cite{khalid2020physical} report two angles of arrival (azimuth and elevation) to counter impersonation in line-of-sight underwater acoustic communication. Recently, the authors in \cite{9676618}  report time reversal resonating strength-based PLA in underwater acoustic sensor networks.

In this paper, we investigate how the physical position/location of the transmitting node can be used as a device fingerprint to detect impersonation attacks at the physical layer in underwater acoustic communication networks. { Although, the authors in \cite{Waqas:Access:2018} use position for authentication, they assumed that the estimates of the position of the transmitter nodes are already available at the receiver, which is a strong and unrealistic assumption. Instead, in this paper, we relax that assumption and for the first time, we systematically exploit the position of the transmitter node for authentication. We provide a step-by-step procedure for position-based PLA from estimation to hypothesis testing.} To the best of our knowledge, this location-based authentication process has not been explored in terrestrial wireless communication networks either. The detailed contributions of this paper are as follows:
\begin{itemize}
    \item We used time of arrival (ToA) based localization to extract the coordinates of the transmitting node. For this, we use the best-unbiased estimator for ToA estimation \cite{Waqas:Access:2018}. We then find the distribution of the inherent uncertainty in the estimation process.
    \item We build a test statistic for binary hypothesis testing in order to decide whether the estimated position belongs to a legitimate node or a malicious node. We derive the distributions of conditional events (i.e., test statistic given legitimate or malicious transmissions) and provide closed-form expressions for the two error probabilities: false alarm rate (FAR) (the probability of classifying a legitimate node as a malicious node) and missed detection rate (MDR) (the probability of classifying malicious  node as a legitimate node).
\end{itemize}
We use simulation-based evaluation to validate our proposed techniques. 



The rest of this paper is organized as follows: Section \ref{sec:sys-model} presents our system model, and  Section \ref{sec:methods} presents the proposed PLA mechanism. We evaluate our proposals using simulation and present the results in Section \ref{sec:results}. Finally, the paper concludes in Section \ref{sec:conclusion} with a few concluding remarks and future directions.

\section{System Model}
\label{sec:sys-model}
\begin{figure}[htb!]
\begin{center}
	\includegraphics[scale=0.35]{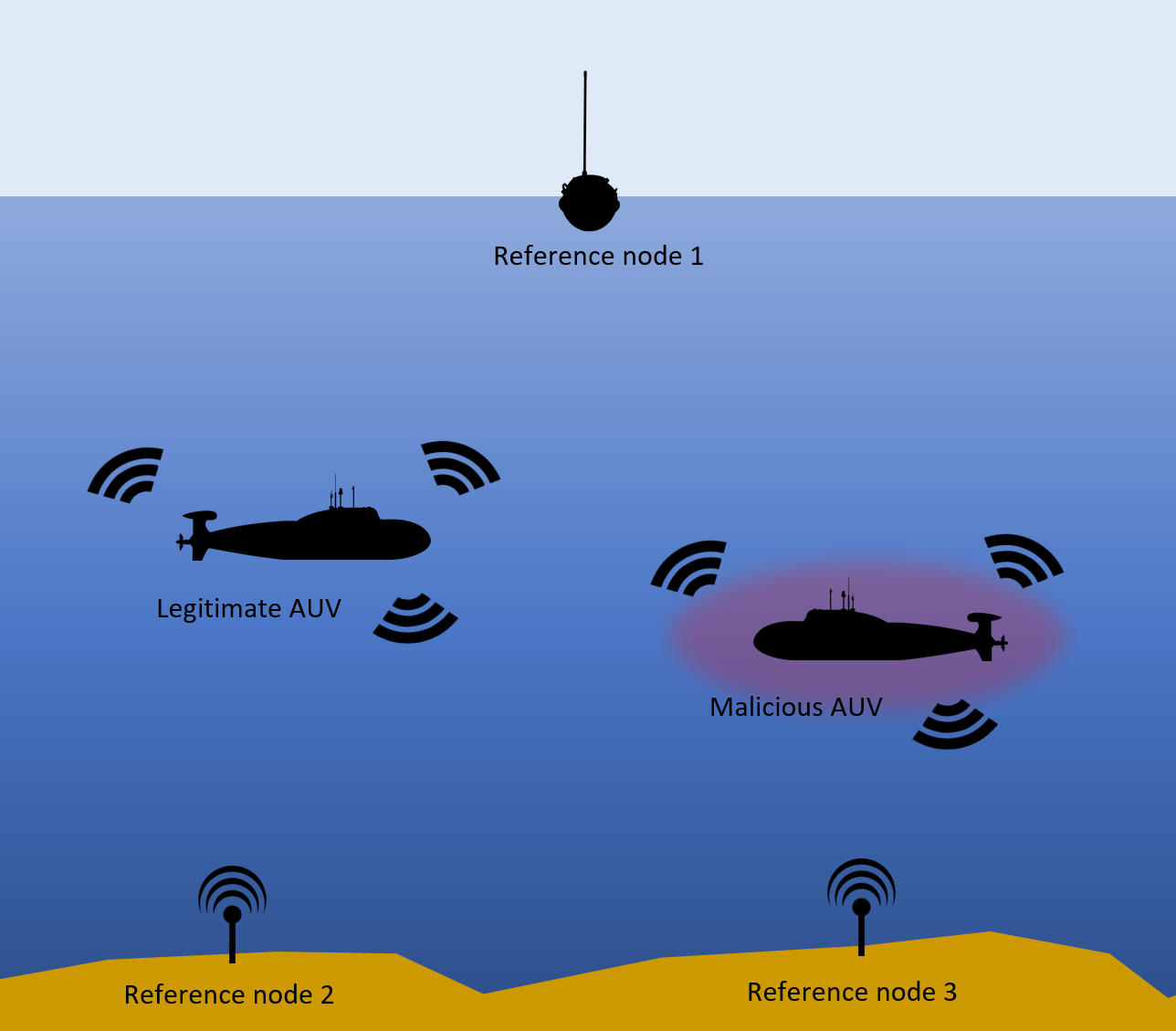} 
\caption{System Model}
\label{fig:sysmodel}
\end{center}
\end{figure}

We consider an underwater scenario as depicted in Fig.\ref{fig:sysmodel}, which consists of two static transmitting nodes, namely: legitimate Autonomous Underwater Vehicle (AUV) and a malicious AUV, and $L$ number of reference nodes, that are used to estimate the position of transmitter AUVs. We assume that the reference nodes are perfectly synchronized and connected to the ground station or surface ship via a secure channel. 

In this scenario, we build a one-way authentication system, where the malicious AUV is an intruder that occasionally sends malicious packets to the system while trying to impersonate the legitimate AUV. This means that  a systematic framework is needed to authenticate the sender of every packet being generated by the transmitting nodes, which will allow the ground station to reject packets from the malicious node. We consider a range-based localization technique (i.e., time of arrival (ToA)) to extract position coordinates from the received signals at the reference nodes. 

We assume that the one-way authentication channel in Fig. \ref{fig:sysmodel} is time-slotted. That is, packets arrive at the reference nodes at discrete-time instants $t_m$ where $t_m-t_{m-1}=T$ is the time gap between two successively received packets at reference nodes. We further assume that malicious AUV transmits with the same transmit power as legitimate AUV in order to stay stealthy in the environment; increasing the power can easily make their transmissions detectable. We also assume that all the $L$ reference nodes are in line-of-sight with the transmitting node as per the requirements of position estimation.

Unless specified otherwise, $\Vert .\Vert_2$ denotes euclidean norm, boldface lowercase letter (i.e., $\mathbf{x}$) and uppercase letter (i.e., $\mathbf{X}$ ) denotes vector and matrix respectively, subscript $._A$ and superscript $.^A$ is used to indicate legitimate node while $._E$ and $.^E$ is used to indicate malicious node, $\sim \mathcal{N}(.,.)$,  $\sim {U}(.,.)$ and $\sim \Gamma(.,.)$ represent distributed as normal, uniform and gamma respectively. Finally, $(.)^T$ is transpose operator, $\mathbb{E}[.]$ is expectation operator and P(.) denotes probability.

\section{Proposed Physical Layer Authentication Framework}
\label{sec:methods}
The proposed PLA mechanism comprises two steps: first, we estimate the position of the transmitter from its transmitted symbols using a range-based localization technique. Next, we do binary hypothesis testing on the estimated position to detect any impersonation attacks. In this section, we discuss these steps in detail.

\subsection{Position estimation}
Position estimation is done via a range-based localization technique, where we use multiple reference nodes (also known as reference nodes) with known coordinates to estimate the distances separating them from the transmitter node based on ToA estimation.  Then, we extract the transmitter's coordinates from the reference nodes' estimated distances by solving the least square problem. 

\subsubsection{Distance Estimation}
Let $t_i$ be the time of arrival (ToA) of a signal at the $i$-th reference node. We exploit the findings of \cite{Waqas:Access:2018} to estimate ToA of the transmitter nodes  in the presence of heavily frequency-dependent pathloss and colored noise using a best-unbiased estimator (i.e., meeting Crammer Rao bound (CRB)). Specifically, the distance between the transmitter node and the $i-$th reference node is estimated using the following distance equation:
$\hat{d_i}=c\hat{t}_i$,   
where $c$ is the speed of sound in water, $\hat{t}_i \sim \mathcal{N}(t_i,\hat{\sigma}_i^2)$ is the estimated ToA with $t_i$ is the actual time of arrival and $\sigma_i^2$ is the variance of the best unbiased estimator \cite{Waqas:Access:2018}, which can be expressed as:
\begin{align}
\label{eq:var}
    \hat{\sigma}_i^2 =  \frac{P_n PL_i(f)}{4P\mathbf{\Hat{s}^TC^{-1}\Hat{s}}},
\end{align}
where $P$ is the transmit power, $P_n$ is the noise power,  $c$ is the underwater speed of sound, $\mathbf{C}$ is the covariance matrix of colored noise. $\mathbf{\Hat{s}}$ is the partial derivative of the pseudo-random sequence $\mathbf{s}$\footnote{A random sequence or message a transmitter needs to transmit for ToA estimation \cite{Waqas:Access:2018} at the reference nodes.}. The estimated distance can be written as: $\hat{d}_i^2=d_i+n_i$,
where $\hat{d_i}$ is the estimated distance, $d_i$ is the true distance of transmitter to the $i$-th reference node, $n_i\sim \mathcal{N}(0,\sigma_i^2)$  is the uncertainty/noise in the estimator with variance $\sigma_i^2= \frac{c^2 P_n PL_i(f)}{4P\mathbf{\Hat{s}^TC^{-1}\Hat{s}}}$. The pathloss $PL_i(f)$ of a transmitter to $i$-th reference node is given as \cite{stojanovic2007relationship}:
\begin{align}
    PL_i(f)_{[dB]}= \nu 10\log_{10}(d)+d\alpha(f)_{[dB]},
\end{align}
where $\alpha(f)_{[dB]}=\frac{0.11f^2}{1+f^2}+\frac{44f^2}{4100+f^2}+2.75\times 10^{-4}f^2+0.003$ with $f$ as operating frequency.
\subsubsection{Coordinates Extraction}
Using the definition of the standard Euclidean distance \footnote{The distance between two points in euclidean space is the euclidean norm of the difference of position vectors of that two points.}, $d_i$ can be written as: $d_i^2=(x-x_i)^2+(y-y_i)^2$. Assuming a high SNR regime we can write $\hat{d_i}^2\approx(d_i+n_i)^2=d_i^2+2n_id_i$.
Now  $\hat{d_i}^2$ in a 2D space can be expressed as
\begin{align}
\hat{d_i}^2=(x-x_i)^2+(y-y_i)^2+2n_i((x-x_i)^2+(y-y_i)^2)^{\frac{1}{2}}.    
\end{align}
Now for all ''$i_s$'' the equation set resulted from the above equation can be written in a matrix-vector form as:
$\mathbf{Ax}+\mathbf{n}=\mathbf{b}$,
where 
\begin{align}
\mathbf{A}=-2
\begin{bmatrix}
    x_1 & y_1 & -0.5   \\
    . & . & .   \\
     . & . & .  \\
    x_L & y_L & -0.5      
  \end{bmatrix} ,  \mathbf{n}=2
  \begin{bmatrix}
    n_1d_1   \\
    .    \\
    . \\
     n_Ld_L     
  \end{bmatrix}, \nonumber \\ \mathbf{x}=\begin{bmatrix} 
    x   \\
    y  \\
     x^2+y^2     
  \end{bmatrix} 
\ \text{and} 
\ \mathbf{b}=\begin{bmatrix}
      \hat{d_1}^2-x_1^2-y_1^2 \\
    .   \\
     .\\
       \hat{d_L}^2-x_L^2-y_L^2   
  \end{bmatrix}. \nonumber 
\end{align}
We can verify that $\mathbb{E}[\mathbf{n}]=\mathbf{0}$, therefore the above matrix-vector form can be approximated as: 
\begin{align}
\label{eq:aprox}
\mathbf{Ax}\approx \mathbf{b}.
\end{align}
Eq. \ref{eq:aprox} leads us to a least square problem, where the solution for $\mathbf{x}$ can be obtained as
\begin{align}
\label{eq:LS}
&\min_{\mathbf{x}}\Vert \mathbf{b}-\mathbf{Ax} \Vert_2^2
=\min_{\mathbf{x}}(\mathbf{b}-\mathbf{Ax})^T(\mathbf{b}-\mathbf{AX}).
\end{align}
One can verify that Eq. \ref{eq:LS} is a convex function. To find the minimum we take the gradient $\nabla_{\mathbf{x}}$ and equate it to zero gives us
 \begin{align}
 \label{eq:sls}
\mathbf{x}=(\mathbf{A}^T\mathbf{A})^{-1}\mathbf{A}^T\mathbf{b},
\end{align}
where $(\mathbf{A}^T\mathbf{A})^{-1}\mathbf{A}^T=\mathbf{A}^{\dagger}$ also known as pseudo inverse of $\mathbf{A}$.
As we are only interested in the $x$ and $y$ components of $\mathbf{x}$, the desired vector can be written as:
$\hat{\mathbf{x}}=[x(1) \ x(2)]^T$.
Next, we derive the distribution of $\hat{\mathbf{x}}$.
As $\mathbf{A}^{\dagger}$ is a $3*L$ dimension matrix and we know that our desired component of $\mathbf{x}$ lies in $\hat{\mathbf{x}}$. Now, let $\hat{\mathbf{A}}^{\dagger}=\mathbf{A}_{2*L}^{\dagger}$, then Eq. \ref{eq:sls} can also be re-written as:
$\hat{\mathbf{x}}=\hat{\mathbf{A}}^{\dagger}\mathbf{b}$,
Or, more precisely, the extracted coordinates  $\hat{\mathbf{x}}$ can be split into two terms (i.e., actual and uncertainty) as:
\begin{align}
\hat{\mathbf{x}}=\underbrace{\hat{{\mathbf{A}}}^{\dagger}\begin{bmatrix}
      {d_1}^2-x_1^2-y_1^2 \\
    .   \\
     .\\
       {d_L}^2-x_L^2-y_L^2   
  \end{bmatrix}}_{\text{True  Co-ordinates}} + \underbrace{2\hat{{\mathbf{A}}}^{\dagger}\begin{bmatrix}
      n_1d_1\\
    .   \\
     .\\
       n_Ld_L   
  \end{bmatrix}}_{\text{Uncertainty}}.
\end{align} 


\subsection{Binary Hypothesis Testing}
 Let $\hat{\mathbf{x}}_A$ be the actual coordinates vector of the legitimate node and $\hat{\mathbf{x}}_E$ of the malicious node. $\text{H}_0$ is the hypothesis (also known as a null hypothesis) that the legitimate node is the transmitter while $\text{H}_1$ is the hypothesis (also known as an alternate hypothesis) that malicious node is the transmitter. Then we define a test statistics TS as:
 
 \begin{align}
    \text{TS}=\Vert ((\hat{\mathbf{A}}^{\dagger})^T\hat{\mathbf{A}}^{\dagger})^{-1}(\hat{\mathbf{A}}^{\dagger})^T\left( \hat{\mathbf{x}}-\hat{
 \mathbf{x}}_A \right)\Vert_2^2.
 \end{align}
Now, the binary hypothesis test can be defined as 
\begin{equation}
	\label{eq:H0H1}
	 \begin{cases} \text{H}_0 (\text{no impersonation}): & \text{TS}=\Vert \hat{\mathbf{A}}^{\dagger\dagger}\left( \hat{\mathbf{x}}-\hat{
 \mathbf{x}}_A \right) \Vert_2^2. < \epsilon_{th} \\ 
                  \text{H}_1 (\text{impersonation}): & \text{TS}=\Vert \hat{\mathbf{A}}^{\dagger\dagger} \left( \hat{\mathbf{x}}-\hat{
 \mathbf{x}}_A \right) \Vert_2. > \epsilon_{th} \end{cases},
\end{equation}
where $\hat{\mathbf{A}}^{\dagger\dagger}=((\hat{\mathbf{A}}^{\dagger})^T\hat{\mathbf{A}}^{\dagger})^{-1}(\hat{\mathbf{A}}^{\dagger})^T$, and $\epsilon_{th}$ is a predefined threshold. 
Equivalently, we have:
\begin{align} 
\label{eq:bht}
\text{TS} \gtrless_{\text{H}_0}^{H_1} {\epsilon_{th}}.
\end{align}
At this stage, we need to find the error probabilities in terms of FAR and MDR. FAR can be defined as the probability that the binary hypothesis test decides a legitimate node as a malicious node while the MDR is the probability that the binary hypothesis test decides a malicious node as a legitimate node. The FAR $P_{fa}$ can be expressed as:
\begin{align}
    P_{fa}=\text{P}(\text{TS} \mid \text{H}_0 > \epsilon_{th}).
\end{align}
To compute the above probability we need to find the distribution of the conditional event $\text{TS} \mid \text{H}_0$, which can be expressed as:
\begin{align}
\label{eq:ts_H_0}
    \text{TS} \mid \text{H}_0&=\Vert \hat{\mathbf{A}}^{\dagger\dagger}\left(\hat{
 \mathbf{x}}_A+\mathbf{n}_A-\hat{
 \mathbf{x}}_A\right) \Vert_2^2=\Vert \hat{\mathbf{A}}^{\dagger\dagger} \mathbf{n}_A\Vert_2^2 \\
 &=4\left( (d_1^An_1^A)^2+(d_2^An_2^A)^2+...+(d_L^An_L^A)^2 \right) \nonumber \\
 &= \sum_{i=1}^L (2 d_i^An_i^A)^2 = \sum_{i=1}^L  (\hat{n}_i^A)^2,  \nonumber
\end{align}
where $\mathbf{n}_A={2\hat{\mathbf{A}}^{\dagger}\begin{bmatrix}
      n_1^Ad_1^A \
    .   \
     . \
       n_L^Ad_L^A   
  \end{bmatrix}} ^T$ with $d_i^A$ where $i \in \{1...L\}$ denotes the distance of legitimate node from $i$-th reference node and $\hat{n}_i^A =(2d_i^An_i^A)\sim \mathcal{N}(0,4(d_i^A)^2(\sigma^A_i)^2), \forall i$.  One can find $(\hat{n}_i^A)^2 \sim \Gamma(1/2, 8(d_i^A)^2(\sigma^A_i)^2)$. Finally, Eq. \ref{eq:ts_H_0} is summation of total $L$ gamma random variables with same shape parameter $1/2$ and different scale parameter $8(d_i^A)^2(\sigma^A_i)^2$.
  So, the probability of false alarm can be computed as \cite{ansari2017new}:
  
  \begin{align}
    P_{fa}  =\sqrt{1/2} \prod_{i=1}^L \sqrt{\frac{1}{\kappa_i}} \ H_{L+1,L+1}^{0,L+1} \left[ e^{\epsilon_{th}} \mid \begin{matrix}
\Xi_L^1, (1,1,1)\\
\Xi_L^2, (0,1,1)
\end{matrix} \right],
  \end{align}
  where $\kappa_i=8(d_i)^2(\sigma_i^A)^2$, $ H_{(.),(.)}^{(.),(.)}$ is the Fox-H function, $\Xi_L^1$ and $\Xi_L^2$ represent the bracket terms as: 
  \begin{align}
      \Xi_L^1= (1-\frac{0.5}{8(d_1^A)^2(\sigma_1^A)^2},1,0.5),...(1-\frac{0.5}{8(d_L^A)^2(\sigma_L^A)^2},1,0.5) \nonumber
  \end{align}
  \begin{align}
      \Xi_L^2= (-\frac{0.5}{8(d_1^A)^2(\sigma_1^A)^2},1,0.5),...(-\frac{0.5}{8(d_L^A)^2(\sigma_L^A)^2},1,0.5). \nonumber
  \end{align}
  
To find the probability of missed detection, we need to find $\text{TS} \mid \text{H}_1$, which can be expressed as 
  \begin{align}
      \text{TS} \mid \text{H}_1&=\Vert \hat{\mathbf{A}}^{\dagger\dagger}\left(\hat{\mathbf{x}}_E+\mathbf{n_E}-\hat{\mathbf{x}}_A\right) \Vert_2^2=\Vert \hat{\mathbf{A}^{\dagger\dagger}} \mathbf{n_E}+\mathbf{d_{EA}}\Vert_2^2 \\
 &=4( (d_1^En_1^E+(d_1^E)^2-(d_1^A)^2)^2+... \nonumber \\ &..+((d_L^En_L^E)+(d_L^E)^2-(d_L^A))^2) \nonumber \\
 &= \sum_{i=1}^L (2 d_i^E\sigma_i^E\hat{n}_i^E)^2 
 = \sum_{i=1}^L (2 d_i^E\sigma_i^E)^2 \chi_1^2(\lambda_i)   \nonumber
  \end{align}
  where $\mathbf{n}_E={2\hat{\mathbf{A}}^{\dagger}\begin{bmatrix}
      n_1^Ed_1^E \
    .   \
     . \
       n_L^Ed_L^E   
  \end{bmatrix}} ^T$ with $d_i^E$ where $i \in \{1...L\}$ denotes the distance of malicious node from $i$-th reference node, $\hat{n}_i^E \sim \mathcal{N}(\lambda_i,1), \forall i$ with $\lambda_i=\frac{(d_1^E)^2-(d_1^A)^2}{\sigma_i^E}$, $\chi_1^2(\lambda_i)$ denotes non-central chi-square random with $1$ degree of freedom and non-centrality parameter $\lambda_i$  and $\mathbf{d_{EA}}={\begin{bmatrix}
      (d_1^E)^2-(d_1^A)^2 \
    .   \
     . \
      (d_L^E)^2-(d_L^A)^2  
  \end{bmatrix}} ^T$. Finally, $\text{TS}\mid\text{H}_1$ is weighted sum of $L$ non-central chi square random variables,
  hence, the computed probability of missed detection $P_{md}=\text{P}(\text{TS} \mid \text{H}_1 < \epsilon_{th})$ can be expressed as \cite{castano2005distribution}:
  
 \begin{align}
     P_{md}= \frac{e^{-\frac{\epsilon_{th}}{2\beta}}}{(2\beta)^{\frac{L}{2}}} \frac{\epsilon_{th}^{\frac{L}{2}}}{\Gamma (\frac{L}{2}+1)} \sum_{k\geq 0} \frac{k!\zeta_k}{(\frac{L}{2}+1)_k}\mathbb{L}_k^{\frac{L}{2}}(\frac{(L+2)\epsilon_{th}}{4\beta \mu_0}),
 \end{align}
  with $\beta >0$, $\mu_0 >0$, $\zeta_k=\frac{\sum_{j=0}^{k-1} \zeta_j \xi_{k-j}}{k}, \gamma_i=4(d_i^E)^2(\sigma_i^E)^2,,$
  \begin{align}
     &\zeta_0=2(\frac{L}{2}+1)^{\frac{L}{2}+1}\exp{-\frac{1}{2}\sum_{i=1}^L \frac{\lambda_i \gamma_i (\frac{L}{2}+1-\mu_0)}{\beta \mu_0+\gamma_i (\frac{L}{2}+1-\mu_0)}}\bullet \nonumber \\ &\frac{\beta^{\frac{L}{2}+1}}{\frac{L}{2}+1-\mu_0)}\prod_{i=1}^L (\beta \mu_0+\gamma_i (\frac{L}{2}+1-\mu_0))^{-\frac{1}{2}}   \nonumber \\
    &\xi_j=-\frac{j\beta(\frac{L+2}{2})}{2\mu_0}\sum_{i=1}^L \lambda_i \gamma_i(\beta-\gamma_i)^{j-1} (\frac{\mu_0}{(\beta \mu_0+\gamma_i (\frac{L+2}{2}-\mu_0))})^{j+1} \nonumber \\ &+(\frac{-\mu_0}{\frac{L}{2}+1-\mu_0})^j+\frac{1}{2}(\frac{\mu_0(\beta-\gamma_i)}{(\beta \mu_0+\gamma_i (\frac{L}{2}+1-\mu_0))})^{j+1}, \nonumber 
  \end{align}
   $\Gamma(.)$ indicates gamma function, $\mathbb{L}_k^{.}$ is the $k$-th  generalized Laguerre polynomial. 

\section{Simulation results}
\label{sec:results}
We use Matlab to develop the simulations presented in this section. We consider a rectangular area of $1000\times1000$ m$^2$, with $L$ total number of reference nodes, a legitimate AUV is fixed at origin (i.e., $X_A=[0,0]$). To obtain a more realistic result, we use the specifications of a commercially available acoustic modem \cite{popoto}. Specifically,  we set the center frequency $f=22$KHz, transmission power $P=100W$, speed of sound $c=1500$m/s, $10$ kHz of bandwidth and spreading factor $\nu=1.5$. We chose two values for reference nodes, $L=3 \ \text{and}\ 5$, where $L=3$ is the minimum number of reference nodes required for ToA-based estimation. The upper limit can be any number greater than $3$ but we chose  $L=5$ for this simulation. We fix the positions of reference nodes to $[0,500], [-500,-500], [500,-500], [-500,500]$ and $[0,-500]$.

\subsection{Error behavior against LQ}
\begin{figure}[htb!]
\begin{center}
	\includegraphics[width=8cm,height=5cm]{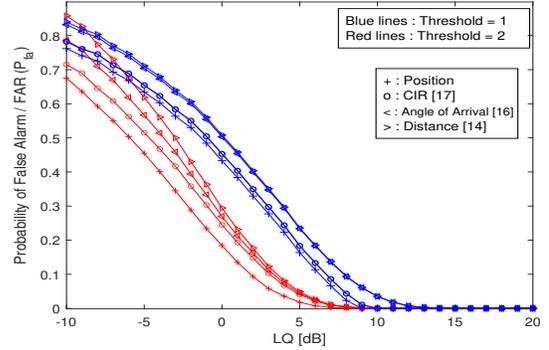} 
\caption{Probability of false alarm $P_{fa}$ against the link quality  in dB.}
\label{fig:fa}
\end{center}
\end{figure}

\begin{figure}[htb!]
\begin{center}
	\includegraphics[width=8cm,height=5cm]{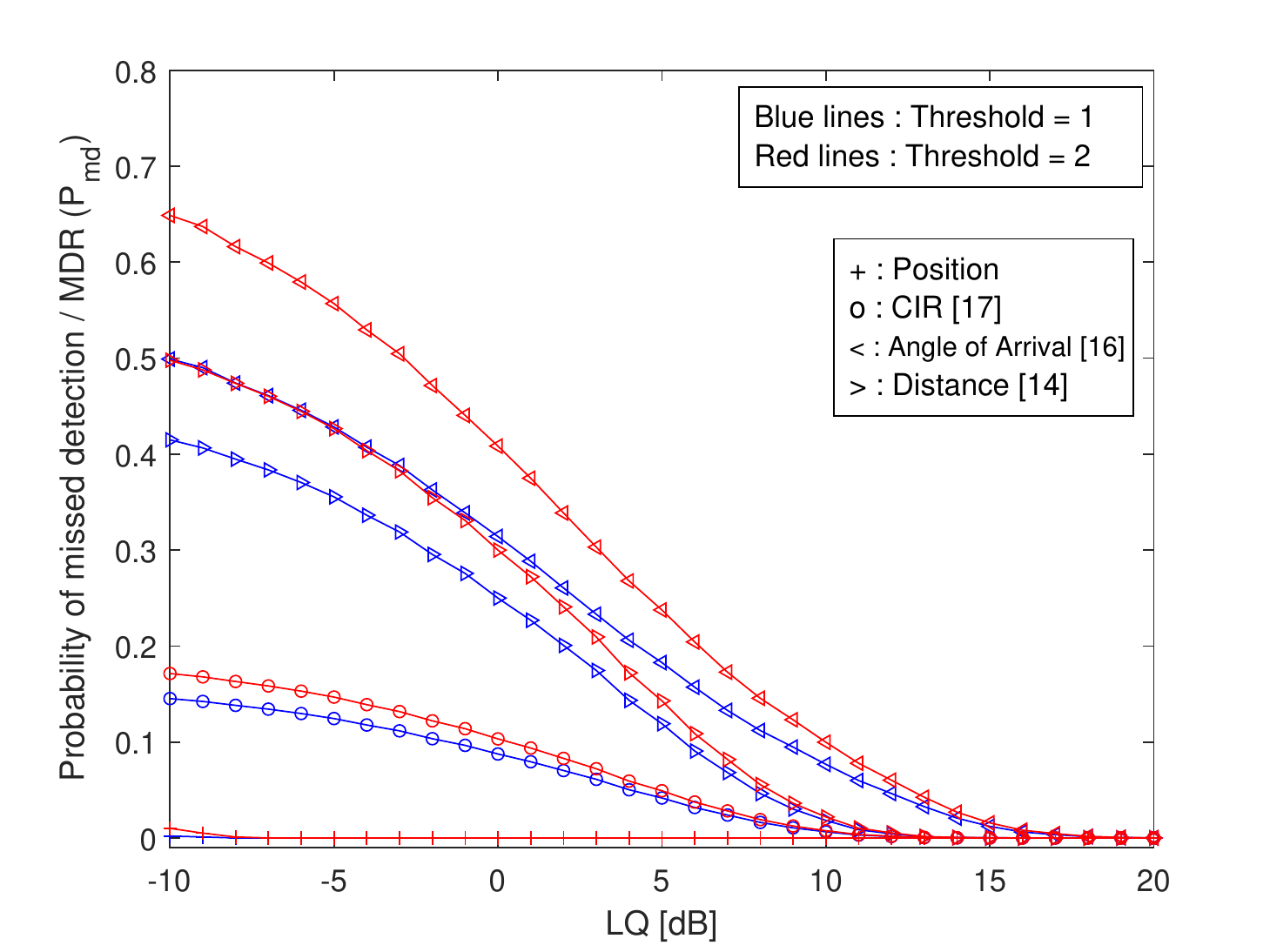} 
\caption{Probability of missed detection $P_{md}$ against the LQ in dB.}
\label{fig:md}
\end{center}
\end{figure}

We choose link quality $\text{LQ}=\frac{P}{P_n}$ (i.e., signal power to noise power ratio) as a controlled parameter or independent variable
to generate Figs. \ref{fig:fa} and \ref{fig:md}. For both figures, we sweep LQ from $-10$ dB to $20$ dB, where $-10$ dB means noise power is $10$ times the signal power and $20$ dB means signal power is $100$ times noise power.
We set $L=3$, and the position of malicious AUV for Figs. \ref{fig:fa} and \ref{fig:md} as a random variable  uniformly distributed in a bounded region around $X_A$ (i.e., $X_E=[\sim U(-500,500m), \sim U(-500,500m)]$). {We compare our mechanism with the performance of previously utilized fingerprints, i.e.,  distance \cite{Waqas:Access:2018}, angel-of-arrival \cite{khalid2020physical} and CIR \cite{9676618} \footnote{The fingerprint used in that work is time reversal resonating strength, which is a derived fingerprint from the fundamental CIR}}. Note that due to multiple reference nodes, the minimum probability of error is considered among $L$ probabilities for all the other fingerprints.

Fig. \ref{fig:fa} demonstrates the trend of the probability of false alarm  {for different fingerprints with the increase in LQ. We can clearly observe that increase in LQ  lowers the FAR. This is due to the fact that the increase in LQ shrinks the variances $\sigma_i^2$ of the $i$-th estimator (Eq. \ref{eq:var}),  which makes the estimated features close to the actual features \cite{kay1993fundamentals}. We also observe that the performance of the position is the best among the others. One can rank the fingerprints in terms of strength as position, CIR, angle of arrival, and distance from strong to weak.} 
On the other hand, there is a clear trade-off between FAR and MDR when varying the threshold $\epsilon_{th}$, as can be seen in the results below.

\begin{figure}[htb!]
\begin{center}
	\includegraphics[width=8cm,height=6cm]{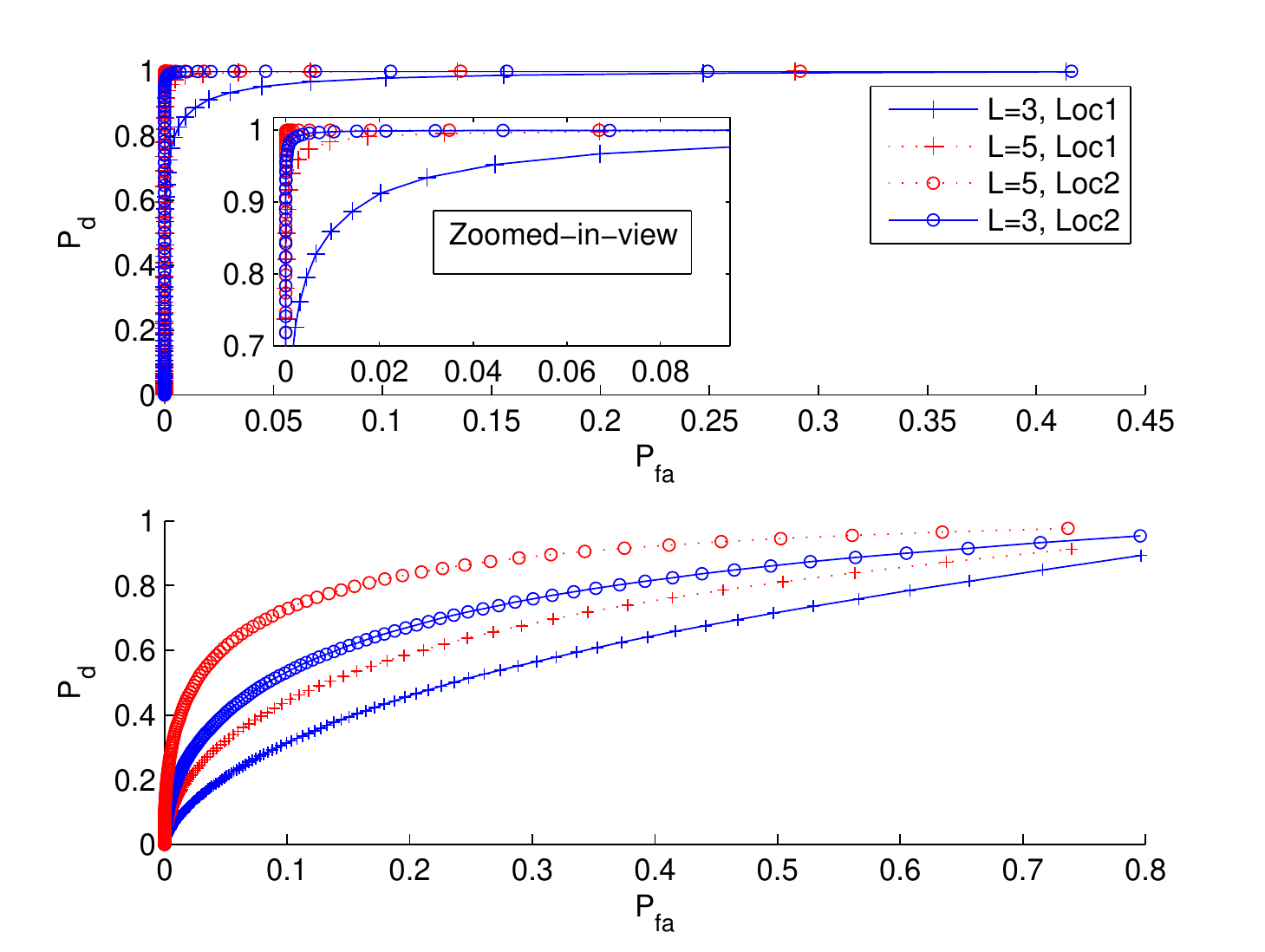} 
\caption{Receiver operating characteristic  (ROC) curves: $P_d$ vs $P_{fa}$. The upper plot is generated with LQ$=10$dB while lower plot is with $0$dB.}
\label{fig:ROC}
\end{center}
\end{figure}

Fig. \ref{fig:md} demonstrates the behavior of MDR against LQ. We observe that the probability of missed detection $P_{md}$ decreases with the increase in LQ and the hypothesis testing threshold $\epsilon_{th}$ has a negative impact on $P_{md}$. This is due to the fact that $\epsilon_{th}$ extends the acceptance range around actual fingerprints of legitimate AUV $X_A$, which makes some of the noisy estimates of malicious AUV's transmissions appear in the acceptance range of legitimate AUV. 
Generally, in PLA, MDR is more important than FAR because it is the probability of accepting malicious transmissions. In Fig. \ref{fig:md}, we can see the actual strength of using the position as a fingerprint for PLA. We can see that MDR for using position is almost zero. In other words, it means that it is nearly impossible for a malicious node to clone the position of a legitimate AUV unless it collides with the legitimate AUV, or, uses high transmission power, but both scenarios will render the malicious node detectable. On the other hand, distance can be cloned while staying away from the legitimate AUV, the angle can also be cloned \cite{Waqas:Access:2018}, and due to the lack of scatters under the water, it is very likely that the two nodes away from each other have the same or closed CIRs as they have the same multi-paths structure.   Furthermore, we observe that the larger the value of $\epsilon_{th}$ the lower the FAR, but at the same time we get a larger MDR. This means that we can not minimize both errors at the same time (we need to trade off one for the other).
\begin{figure}[htb!]
\begin{center}
	\includegraphics[width=8cm,height=5cm]{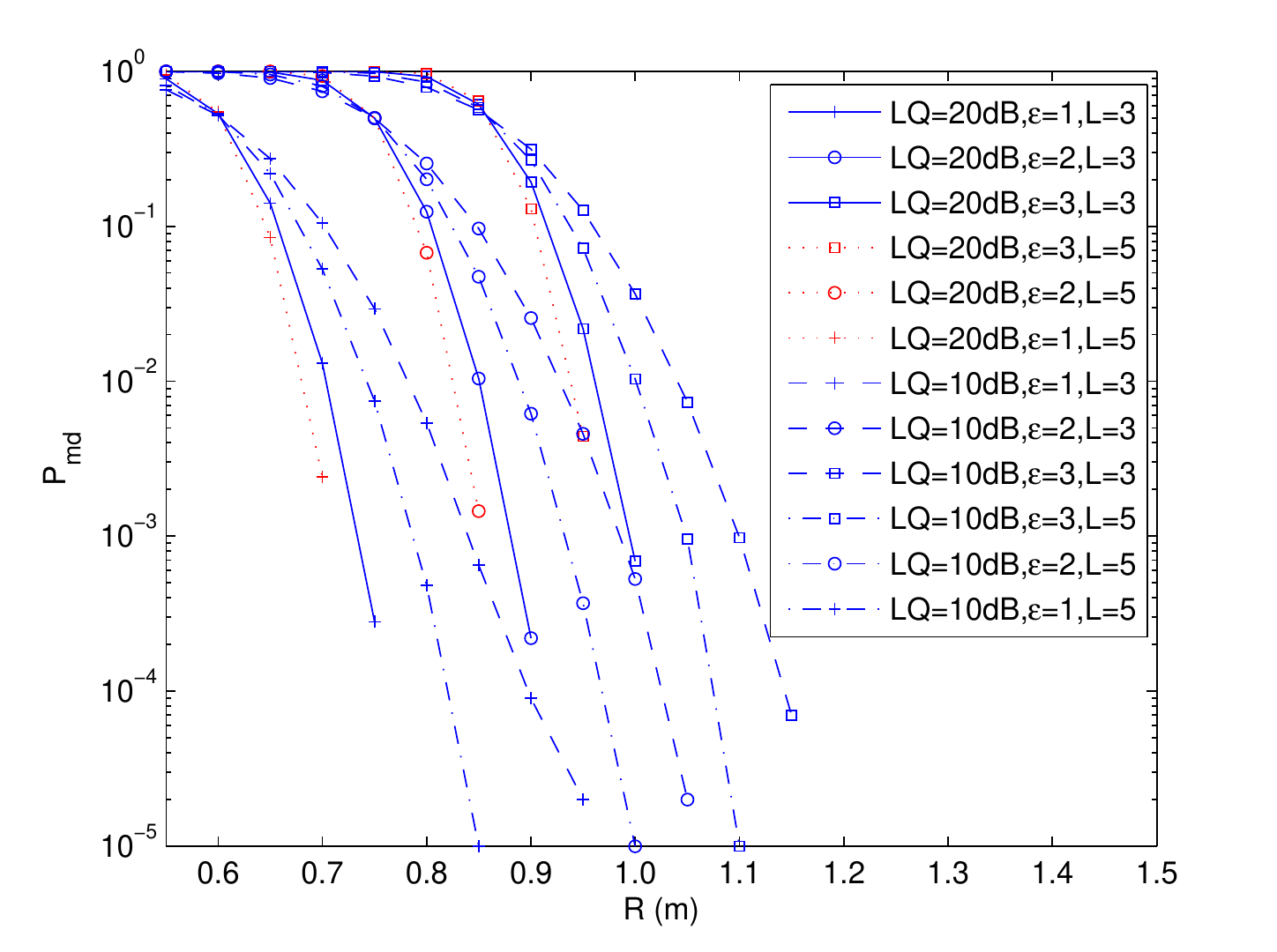} 
\caption{Probability of missed detection $P_{md}$ against malicious node distance from legitimate node}
\label{fig:pmdvsR}
\end{center}
\end{figure}
\\
Next, to have a more detailed evaluation of our proposed mechanism, we generate Figs. \ref{fig:pmdvsR} and \ref{fig:ROC}.
\subsection{Impact of malicious AUV location}
 To see the impact of near and far location of malicious AUV, we generate Fig. \ref{fig:pmdvsR}. In this plot, $R$ is the distance/radius of a circle from the center/$X_A$, and malicious AUV is placed randomly at the circle of radius $R$. For better exposition, we set the Y-axis to log scale. Surprisingly, we observe that when a malicious node is close to the actual node for fixed values of $\epsilon_{th}$ and $L$, increasing LQ produces high MDR (which can be seen in Fig. \ref{fig:pmdvsR}) but quickly gets back to normal behavior (i.e., more LQ produces low MDR) as the malicious AUV goes farther away from $X_A$. We suspect that this abnormal behavior is due to the fact that more LQ means the noisy estimates are close to the true value and thus when malicious AUV is very close to legitimate AUV, the noisy estimates of malicious AUV are close to legitimate AUV, which is classified by hypothesis testing as legitimate AUV. We leave finding the exact position of malicious AUV from where the abnormal behavior starts as a future work. 
\subsection{ROC}

Receiver operating characteristic (ROC) curves are important to evaluate PLA. Fig. \ref{fig:ROC} presents the ROC of the proposed PLA mechanism. It is generated by carrying out an exhaustive search over $\epsilon_{th}$ and recording $P_{fa}$ and $P_{md}$. Fig. \ref{fig:ROC} shows a relation between the probability of detection $P_d=(1-P_{md})$ and FAR for two different locations (i.e., Loc1  $X_E=[1m,1m]$ and Loc2 $X_E=[2m,2m]$) of malicious AUV and total number of reference nodes $L$. The upper subplot is generated for LQ$=0$dB while the lower subplot is generated for LQ$=10$dB. We observe that increasing LQ produces a high detection rate for low FAR. $L$ has a similar effect on the detection rate as we see in the last figures, an increase in $L$ produces a high detection rate. As expected we observe the impact of malicious node location on ROC, a malicious node near to legitimate node (i.e., Loc1) produces lower ROC than a farther location (i.e., Loc2). These ROC curves attest to the efficacy of using the position as device fingerprint for PLA where we can see that for moderate link quality LQ$=10$dB and $10\%$ FAR, we achieve $100 \%$ detection rate when malicious AUV is just 1m away from legitimate AUV. 
\section{Conclusion}
\label{sec:conclusion}
In this paper, we studied physical layer authentication (PLA) based on the position of the transmitting node. The position is estimated based on time of arrival (ToA) estimation for which the best-unbiased estimator was used. The distribution of uncertainty in estimation was derived. Further, to counter impersonation, binary hypothesis testing was used to classify the transmissions of transmitters into legitimate or malicious transmissions. The closed-form expression for errors resulting from hypothesis testing was derived. {The simulation results were compared with the performance of other fingerprints utilized for PLA in the previous work. We observed from the simulation results that position can be used as a device fingerprint for PLA and it is superior to other fingerprints (particularly, in terms of MDR)}. In particular, the ROC curves obtained from our simulation show that for low false alarm and reasonable link quality, $100 \%$ probability of detection can be achieved. Future extension of this work will consider PLA for mobile transmitters.

\section*{Acknowledgement}
This work is partially funded by the G5828 ``SeaSec: DroNets for Maritime Border and Port Security" project under the NATO Science for Peace and Security Programme.






\bibliographystyle{IEEEtran}
\bibliography{references}

\vfill\break

\end{document}